\documentclass[twocolumn,aps,prd,nofootinbib,a4paper,floatfix]{revtex4-2}
\usepackage{latexsym,amsbsy,amsmath,bm,color}
\usepackage{graphicx}
\usepackage[utf8]{inputenc}
\usepackage{microtype}
\mathchardef\mhyphen="2D
\def\vec#1{\bm{#1}}

\def\ll#1#2{\tilde{\lambda}_{#1}.\tilde{\lambda}_{#2}}

\begin{document}

\title{Effect of relativistic kinematics on the stability of multiquarks}
\author{Jean-Marc~Richard}
\email{j-m.richard@ipnl.in2p3.fr}
\affiliation{Universit\'e de Lyon, Institut de Physique des 2 Infinis de Lyon,
IN2P3-CNRS--UCBL,\\
4 rue Enrico Fermi, 69622  Villeurbanne, France}
\author{Alfredo~Valcarce}
\email{valcarce@usal.es}
\affiliation{Departamento de F{\'\i}sica Fundamental,\\
Universidad de Salamanca, 37008 Salamanca, Spain}
\date{\emph{Version of }\today}
\author{Javier~Vijande}
\email{javier.vijande@uv.es}
\affiliation{Unidad Mixta de Investigaci\'on en Radiof\'\i sica e Instrumentaci\'on 
Nuclear en Medicina (IRIMED)\\ 
Instituto de Investigaci\'on Sanitaria La Fe (IIS-La Fe)\\
Universitat de Valencia (UV) and IFIC (UV-CSIC), Valencia, Spain}
\begin{abstract}
We discuss whether the bound nature of multiquark states in quark models could benefit from  relativistic effects on the
kinetic energy operator. For mesons and baryons, relativistic corrections to the kinetic energy lead to lower energies, 
and thus call for a retuning of the parameters of the model. For multiquark states, as well as  their respective thresholds, 
a comparison is made of the results obtained with non-relativistic and relativistic kinetic energy. It is found that 
the binding energy is lower in the relativistic case. In particular, $QQ\bar q\bar q$ tetraquarks with double heavy flavor  become stable for a 
larger ratio of the heavy to light quark masses; and the all-heavy tetraquarks $QQ\bar Q\bar Q$ that are not stable in standard non-relativistic quark models remain unstable when a relativistic form of kinetic energy is adopted.  
\end{abstract}
\maketitle
\section{Introduction}
\label{se:intro}
In a rather celebrated paper~\cite{Capstick:1986kw}, Isgur and his collaborators analyzed how to take out the ``naive'' and ``non-relativistic'' out of the quark model. After this pioneering work, there have been several other studies on how to implement a minimal amount of relativity in the quark model, e.g.,~\cite{Basdevant:1984rk,Capstick:1986bm,1992IJMPA...7.6431L,Brau:2004ie}.

We are aware that this is just a small part of the problem. For instance, in the case of the positronium atom or ion, there are many effects, very often canceling each other, and the kinematics is just one of them. See, e.g.,~\cite{2005JPhB...38.3377D,thaller1992dirac}. We nevertheless 
deemed appropriate to examine the role of relativistic kinematics in the quark model and to go beyond the case of mesons and baryons.
The main motivation is the current interest about the stability of multiquark hadrons and the delicate interplay between the 
energy of collective configurations and their corresponding thresholds. 
This is precisely the aim of the present article, to study to which extent the stability of tetraquarks is influenced by relativistic kinematics. This implies that the mesons constituting the threshold and the tetraquarks are estimated consistently within the same framework, either non-relativistic (NR) or relativistic, or, more precisely, semi-relativistic (SR), to keep in mind that we take into account only a fraction of the relativistic effects. For a given potential, all hadron masses tend to decrease if one adopts a relativistic form of the kinetic energy.  It will be shown that the mesons are more affected than the tetraquarks, so that the binding energy with respect to the lowest threshold becomes smaller for bound tetraquarks. 

For each hadron, we concentrate on its energy $E$, so that its mass is given by $\mathcal{M}=\sum m_i+ E$, where the $m_i$ are the constituent masses of its quarks.  For the discussion about the stability of tetraquarks, we have the same cumulated constituent mass, $\sum m_i$, in the tetraquark and in the mesons entering its threshold.

The paper is organized as follows. In Sec.~\ref{se:form}, we review the formalism, focusing on how to estimate the matrix elements in a variational calculation using a basis of correlated Gaussians. Then, some applications are given for ordinary hadrons in Sec.~\ref{se:ord},  and for tetraquarks in Sec.~\ref{se:tetra}. In particular, we discuss how the relativistic effects influence the binding of doubly-heavy tetraquarks with respect to their dissociation into two flavored mesons and how the fully-heavy tetraquarks are affected by the choice of kinematics. Some conclusions are drawn in Sec.~\ref{se:out}. 
\section{Formalism}\label{se:form}
Let us start with the one-body Hamiltonian in three dimensions
\begin{equation}\label{eq:one-b}
 H_1=K(\vec p,m)+V(r)~,
\end{equation}
with $K(\vec p,m)$ being either $K_{\rm NR}=\vec p^2/(2\,m)$ or $K_{\rm SR}=(\vec p^2+m^2)^{1/2}- m$.
The ground state is sought at variationnally using a trial wave function 
$\Psi(\vec r)=Y_{00}(\hat r) u(r)/r$ with
\begin{equation}\label{eq:exp}
u(r)=\sum_{i=1}^N \gamma_i\, r\,\exp(-a_i\,r^2/2)=\sum_{i=1}^N \gamma_i\, w(a_i,r)~.
\end{equation}
The computation involves the matrix elements of normalization, potential and kinetic energy between two basis functions $w(a,r)$ and $w(b,r)$, which are noted as
\begin{eqnarray}\label{eq:mat-elt}
n(a,b)&=&\frac{\sqrt{\pi}}{4\,c^{3/2}}~,\notag \\
v(a,b)&=&=n(a,b)\, g_V(c)~,\\
k(a,b)&=&=n(a,b)\, f(\bar c)~,\notag
\end{eqnarray}
where $c=(a+b)/2$ and $\bar c=2\,a\,b/(a+b)$ are  suitable averages. 
By straightforward calculation, one obtains $g_V(c)=3/(2\,c)$ for an harmonic potential, $2/\sqrt{c\,\pi}$ for a linear one,  $2\,\sqrt{c/\pi}$ for a Coulomb one, etc. As for the kinetic energy, in the NR case, one gets $f_{\rm NR}(\bar c)=3\,\bar c/(4\,m)$, while in the SR case, it is 
\begin{equation}\label{eq:elt-K}
 f_{\rm SR}(\bar c)=\sqrt{\frac{1}{\bar c}}\,m^2\,\exp\genfrac{(}{)}{}{}{m^2}{2 \bar c}\,K_1\genfrac{(}{)}{}{}{m^2}{2 \bar c}-m~,
\end{equation}
where some care is required in the computation of the Bessel function $K_1$, at least with some of the available softwares. 

The variational expansion~\eqref{eq:exp} converges remarkably well. For instance, for $m=1$ and $V(r)=r$, for which the ground state energy is exactly the negative  of the first root of the Airy function, one can estimate 
\begin{equation}
\delta_2 H= \langle H^2\rangle - \langle H \rangle^2~,
\end{equation}
whose vanishing ensures that the Temple-Kato lower bound coincides with the variational upper bound, as explained, e.g., in~\cite{Thirring:2023497}. For a single Gaussian ($N=1$), one gets $\delta_2 H=0.027$, and for $N=6$, $\delta_2=0.00002$. It has been checked that a similar convergence is obtained for other potentials and/or kinetic-energy operators.

For a meson, in the center of mass, the NR Hamiltonian is just \eqref{eq:one-b} with $m$ in $K_{\rm NR}(\vec p,m)$ being the reduced mass. For the SR Hamiltonian, one simply adds up $K_{\rm SR}(\vec p,m_1)$ and $K_{\rm SR}(\vec p,m_2)$. Hence, the estimate of the energy and wave function is rather similar to that of the one-body case. 

For a baryon, in the center of mass, the Hamiltonian reads
\begin{equation}\label{eq:H3}
H_3=\sum_{i=1}^3 K(\vec p_i,m_i)+\sum_{i<j} V(r_{ij})~,
\end{equation}
where $r_{ij}=|\vec r_j-\vec r_i|$, and $\sum \vec p_i=0$.

In the NR case, it is customary, but not compulsory, to introduce Jacobi coordinates such that the total kinetic energy operator becomes diagonal. For our purpose, it is sufficient to introduce an universal set of variables in momentum space
\begin{equation}\label{eq:Jaco-mom3}
 \begin{aligned}
  \vec P&=(\vec p_1+\vec p_2+\vec p_3)/\sqrt3~,\\
  \vec p_x&=(\vec p_2-\vec p_1)/\sqrt2~,\\
  \vec p_y&=(2\,\vec p_3 -\vec p_1-\vec p_2)/\sqrt6~,
 \end{aligned}
\end{equation}
and their conjugate
\begin{equation}
 \begin{aligned}
  \vec R&=(\vec r_1+\vec r_2+\vec r_3)\sqrt3~,\\
  \vec x&=(\vec r_2-\vec r_1)/\sqrt2~,\\
  \vec y&=(2\,\vec r_3 -\vec r_1-\vec r_2)/\sqrt6~,
 \end{aligned}
\end{equation}
so that the individual momenta $\vec p_i$ are linear combination of $\bm P$, $\vec p_x$ and $\vec p_y$, and the distances $\vec{r}_{ij}$ linear combinations of $\vec x$ and $\vec y$.  The most general spatial wave function corresponding to an overall $S$-wave reads
\begin{equation}\label{eq:Psi}
 \Psi(\vec x,\vec y)=\sum_i\gamma_i \,\exp\left[-\{\vec x,\vec y\}.A_i.\{\vec{x},\vec{y}\}^t\right]~,
\end{equation}
or, in abbreviated form, $|\psi\rangle=\sum \gamma_i \,|A_i\rangle$, 
where the symmetric, definite-positive, matrices $A_i$ contain the range coefficients. As in the one-body case, the minimisation is reached in two steps: for a given set of $A_i$, the coefficients $\gamma_i$ and the variational energy $\langle \Psi|H_3|\Psi\rangle/\langle \Psi|\Psi\rangle$ are given by an eigenvalue equation, and this energy is minimized by varying the $A_i$. 

The matrix elements of interest are obtained by standard techniques of Gaussian integration \cite{zinn2002quantum,Fedorov:2017bcq}. If a pair corresponds to a separation $\vec r_{ij}= a\,\vec x+b\,\vec y=\vec\alpha.\{\vec x,\vec y\}^t$, then the matrix element of $V(r_{ij})$ is 
\begin{equation}
 \langle A|V(r_{ij})|B\rangle=
 \langle A|B\rangle\,g_V(a_e)~,\quad
 a_e=1/{\vec\alpha.C.\vec\alpha^t}~,
\end{equation}
where $C=(A+B)/2$. 

Similarly, for the kinetic energy, if, say $\vec p_1=u\,\vec p_x+v\,\vec p_Y+w\,\vec P=\vec\beta.\{\vec p_x,\vec p_Y\}+w\,\vec P$, then
\begin{equation}\label{eq:K}
 \langle A|K(\vec p_1,m_1)|B\rangle
 =\langle A|B\rangle\,f(\bar a_e)~,\quad
 \bar a_e=\vec\beta.\bar C.\vec\beta^t~, 
\end{equation}
where $2\,\bar C^{-1}=A^{-1}+B^{-1}$. 

The extension to tetraquarks is straightforward. One can introduce momenta such as
\begin{equation}\label{eq:Jaco-mom4}
 \begin{aligned}
  \vec P&=(\vec p_1+\vec p_2+\vec p_3+\vec p_4)/2~,\\
  \vec p_x&=(\vec p_2-\vec p_1)/\sqrt2~,\\
  \vec p_y&=(\vec p_4-\vec p_3)/\sqrt2~,\\
  \vec p_z&=(\vec p_3+\vec p_4 -\vec p_1-\vec p_2)/2~,
 \end{aligned}
\end{equation}
and the conjugate variables in position space, and express the individual momenta and the relative distances in terms of them. Then, Eqs.~\eqref{eq:Psi}-\eqref{eq:K} are easily extended from dimension $2\times2$ to $3\times3$.

The convergence of the variational expansion~\eqref{eq:Psi} is as good as for the one-body case. The Gaussian expansion method, indeed, is well documented, as discussed, e.g.,  in a paper written by some of the best experts of few-body physics, with benchmark calculations in various fields of physics~\cite{Mitroy:2013eom}. In the case of baryons (or multiquarks), care should be taken that all degrees of freedom are incorporated, and this is achieved if one introduces $2\times2$ ($3\times3$ for tetraquarks) matrices which are not restricted to a scalar or diagonal form. For instance, if a system of three equal masses is described with an expansion 
$\sum \gamma_i\,\exp(-a_i(\vec x^2+\vec y^2)/2)$, one will never account for the possibility of internal orbital momenta $\ell_x=\ell_y>0$.

Note that the formalism does not need to be modified to handle unequal masses. On can still use the Jacobi coordinates~\eqref{eq:Jaco-mom3} and~\eqref{eq:Jaco-mom4} in momentum space and their conjugates in position space. 
Let us insist on that the center-of-mass energy is \emph{exactly} removed if one adopts a variational wave function that is invariant under translations. For instance, in atomic physics, many estimates of $M^+ e^- e^-$ ions are made in
a frame attached to the positive nucleus, and the corrections are
taken into account by the ``mass-polarization'' term. The beautiful proof of the stability of the positronium molecule by two pioneers of quantum physics~\cite{PhysRev.71.493} was achieved with a trial wave-function that depends only on some relative separations $\vec r_j-\vec r_i$. Nevertheless, this proof was unjustly criticized by arguing that the center of mass is not removed beforehand in the Hamiltonian~\cite{1968PhRv..171...36S}.
\section{Ordinary hadrons}\label{se:ord}
\subsection{Mesons}\label{subse:mesons}
\begin{figure}[t!]
 \centering
 \includegraphics[width=.35\textwidth]{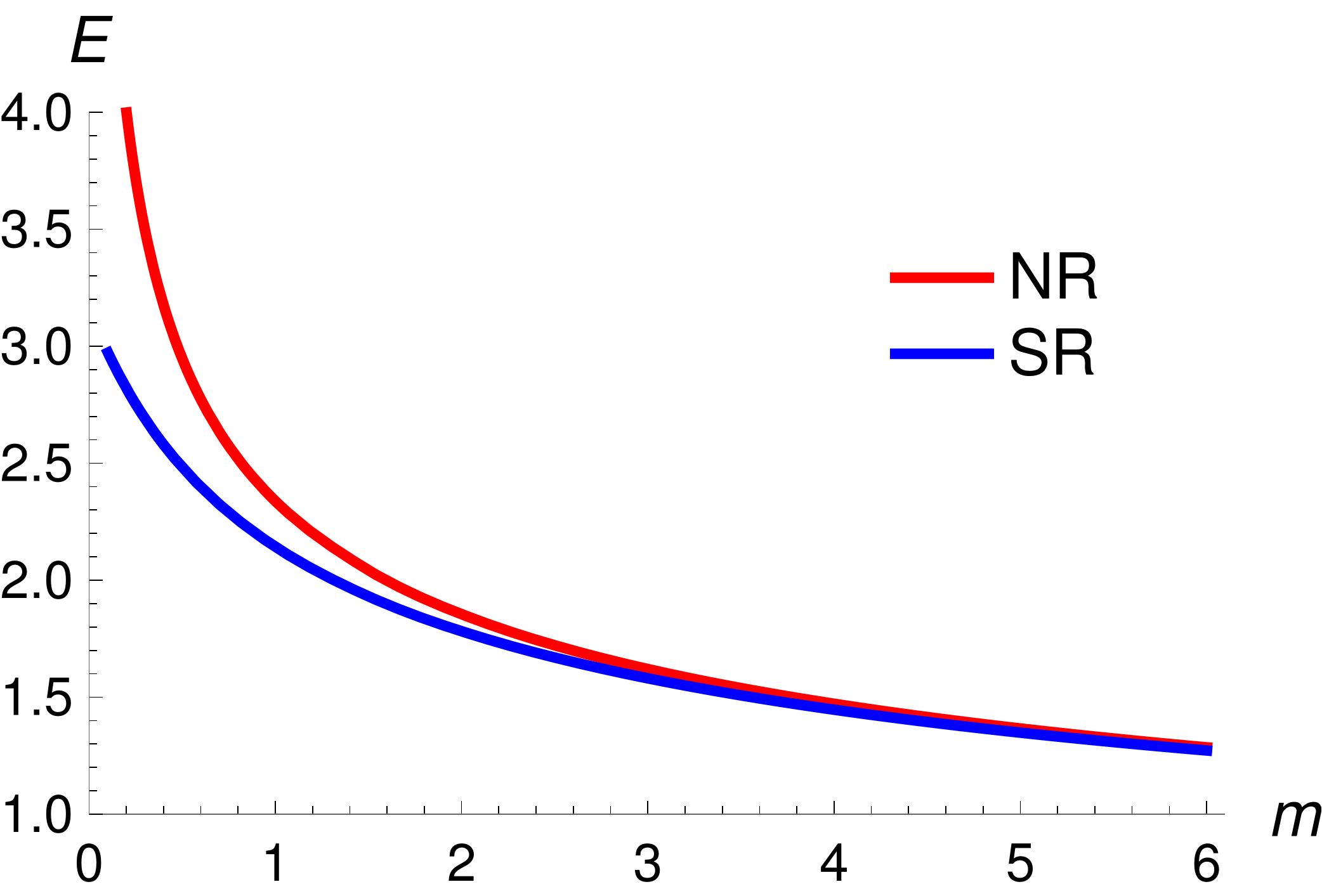}
 \caption{Semi-relativistic vs.\ non-relativistic ground-state energy of a $(m,m)$ meson bound by a potential $V(r)=r$. The energy scale is fixed by the mass $m$ and the string tension set to unity.}
 \label{fig:meson1}
\end{figure}

As a first illustration, we compare in Fig.~\ref{fig:meson1} the energy calculated with a potential $V(r)=r$ and constituent masses $(m,m)$. The energy scale is given by the string tension set to $\sigma=1$. In the NR case, all energies are proportional to $m^{1/3}$, while in the SR case, they are recomputed for each $m$. Not surprisingly, the SR energy is lower than the NR one. Indeed,
\begin{equation}
 K_{\rm SR}(\vec p,m)\le K_{\rm NR}(\vec p,m)~.
\end{equation}
We now turn to the case of a pure Coulomb interaction. Both NR and SR energies are proportional to $m$, so we set  $m=1$ and deal with the well-studied ``Herbst'' Hamiltonian 
\begin{equation}\label{eq:herbst}
 h_{\rm SR}(\alpha)=\sqrt{\vec p^2+1}-1-\frac{\alpha}{r}~,
\end{equation}
and its NR analog with $\vec p^2/2$. As discussed in the literature \cite{1977CMaPh..53..285H,1992IJMPA...7.6431L,Brambilla:1995mg,1997JMP....38.3997L,Raynal:1993qh}, the Herbst Hamiltonian becomes delicate when $\alpha$ approaches $2/\pi$. It is not our aim to enter the mathematical subtleties of \eqref{eq:herbst}, but to stress that
\begin{itemize}
 \item in the Coulomb regime, the relativistic effects are governed by the strength of the attractive $1/r_{ij}$ terms,
 \item when this strength increases, the convergence of the variational expansion becomes more delicate.
\end{itemize}

\begin{figure}[h!]
 \centering
 \includegraphics[width=.3\textwidth]{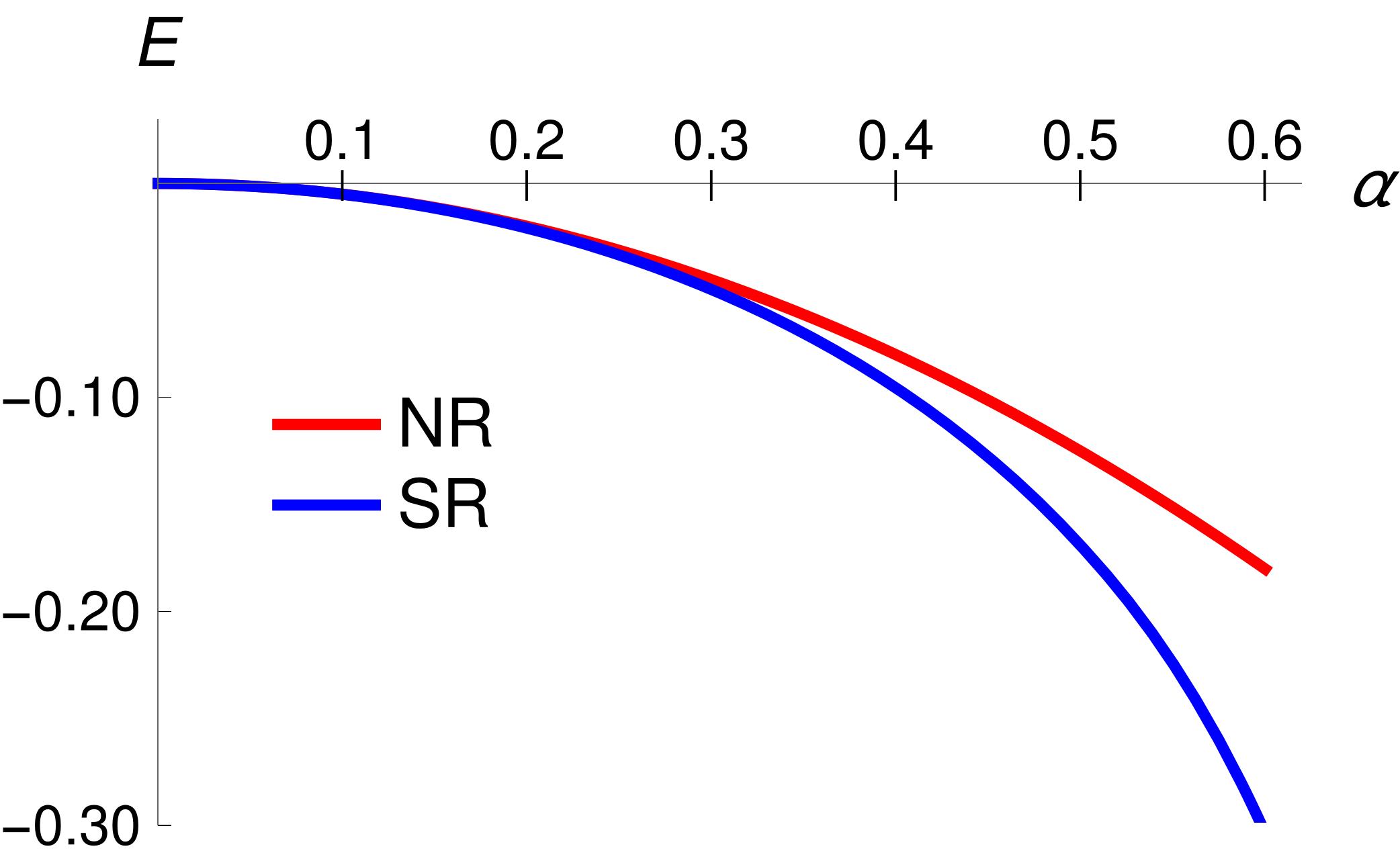}
 \caption{Numerical evaluation of the ground state of the Herbst Hamiltonian \eqref{eq:herbst} vs.\ the NR energy for a $(m,m)$ system with $m=2$, bound by a Coulomb potential $\alpha/r$. The energy scale is fixed by the mass $m$.}
 \label{fig:herbst}
\end{figure}

In Fig.~\ref{fig:herbst} is shown a comparison of the ground state of the Herbst Hamiltonian  \eqref{eq:herbst} and its NR version.

\begin{figure}[h!]
 \centering
 \includegraphics[width=.35\textwidth]{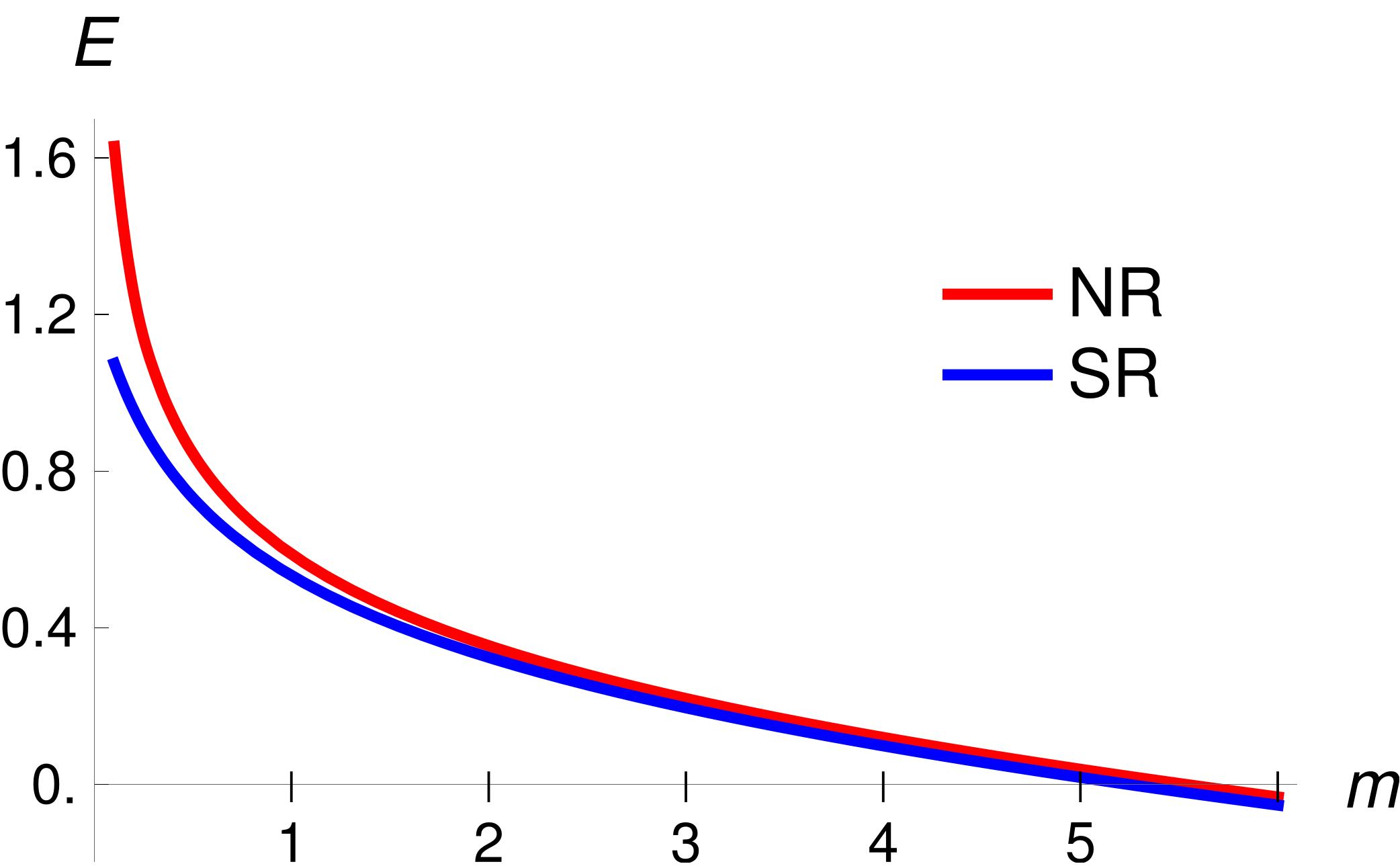}\\[.2cm]
 \includegraphics[width=.35\textwidth]{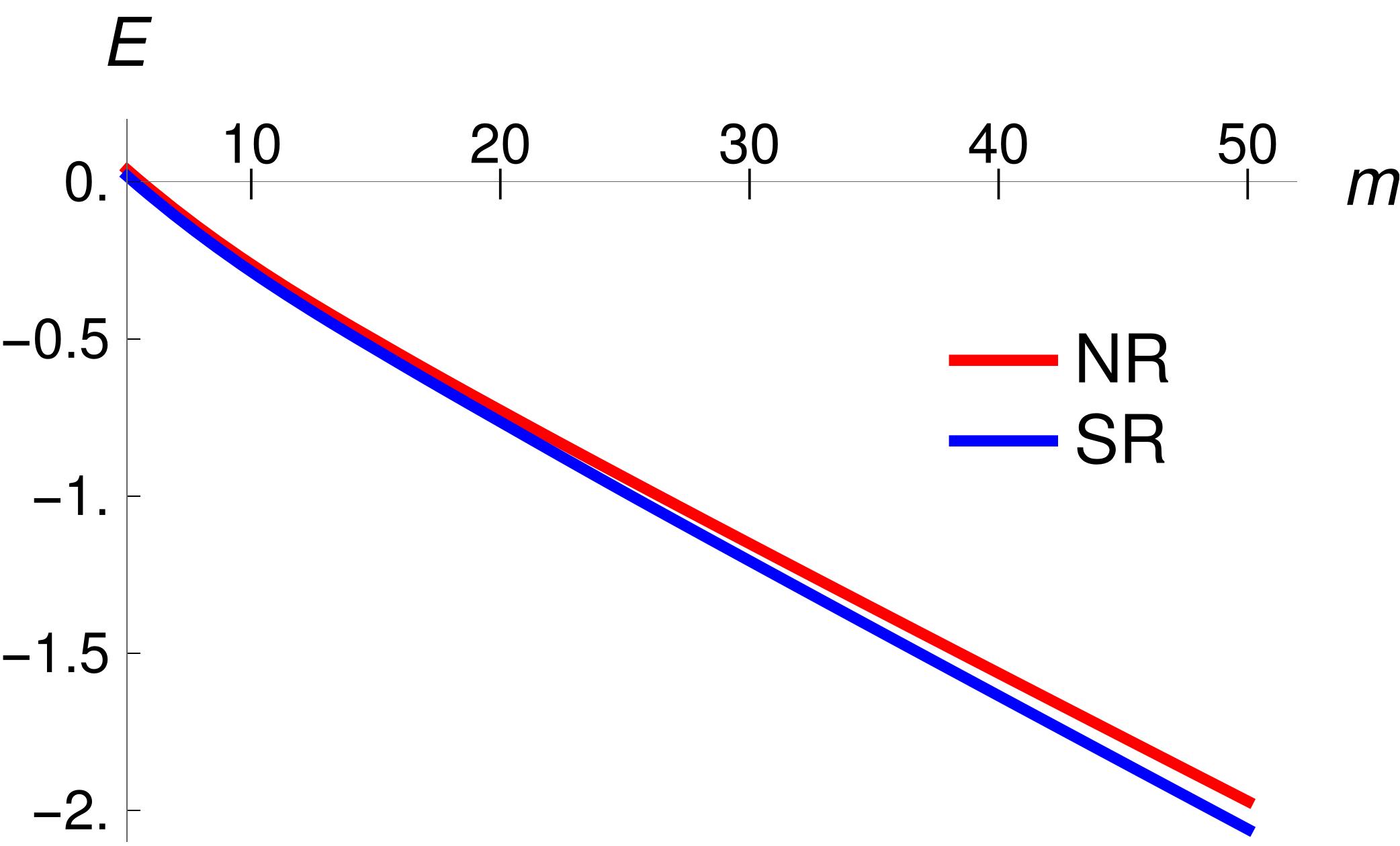}
 \caption{SR vs.\ NR ground state energy of $(m,m)$ mesons in a potential $V(r)=-0.4/r+0.2\,r$, as a function of the mass of the constituents. Energy in GeV.}
 \label{fig:meson-cl}
\end{figure}

Now for a realistic potential such as $V(r)=-0.4/r+0.2\,r$, where $V$ is in GeV and $r$ in GeV$^{-1}$, the SR and NR energies of $(m,m)$ mesons significantly differ for light quarks, say $m<1\,$GeV, and become close in the heavy quark regime, say $m\sim 3\,\mhyphen\,5\,$GeV. But for very heavy constituents, the Coulomb interaction becomes dominant, and the ratio of energies increases again. This is shown in Fig.~\ref{fig:meson-cl}.

For all above potentials, the decrease of the meson energy when going from the NR to SR case is accompanied  by a significant increase of the wave function at the origin,$\Psi(0)=u'(0)/\sqrt{4\,\pi}$. For instance for a binary system with masses $m_1=m_2=1$ bound by $V(r)=r$, one gets exactly $|u'(0)|^2=1$, while the SR analog is $|u'(0)|^2\simeq 1.8$. For a Coulomb interaction $-g/r$ , the effect is even larger: the ratio of the $|u'(0)|^2$ is nearly 3 for $g=1/2$. 

For quark models, this dramatic enhancement of the short-range correlations implies a drastic re-tuning of the spin-spin part of the interaction, whether or not it is treated as first-order perturbation, or included non-perturbatively in the model. For the latter option, an example is the AL1 potential~\cite{Semay:1994ht}, according to which the interaction of a quark of mass $m_i$ and an antiquark of mass $m_j$ is 
\begin{gather}
V_{ij}(r)= -\frac{\kappa}{r} +\lambda\,r-\Lambda
{}+ \frac{2\,\pi\,\alpha}{3\,m_i\,m_j}
\frac{\exp(-r^2/r_0^2)}{\pi^{3/2}\,r^3_0}\,\vec\sigma_i.
\vec\sigma_j~,\nonumber\\
r_0(m_i,m_j)=A\,\genfrac{(}{)}{}{0}{2\,m_i\,m_j}{m_i+m_j}^{-B}~,\label{eq:AL1}\\
m_q=0.315~,\quad m_c=1.836~,\quad m_b=5.227~,\ \nonumber\\
\Lambda=0.8321~,\quad B=0.2204~,\quad A=1.6553~,\nonumber \\
\kappa=0.5069~,\quad \alpha=1.8609~,\quad \lambda=0.1653~.\nonumber
\end{gather}
It fits rather well the ground-state hadrons, and in particular the spin multiplets of interest in the study of doubly-flavored tetraquarks, for instance $m(D) = 1862\,$MeV and $m(D^*) = 2016\,$MeV.
%

However, if one adopts the SR kinetic energy, this potential does not describe reasonably well the quarkonia nor the flavored mesons, as it leads to hyperfine splittings that are much too large. Thus, for the SR calculations, we used a modified version, hereafter referred to as AL1N, which consists of 
\begin{equation}\label{eq:AL1N}
 m_c\ \to\ m'_c=2.007~\mathrm{GeV} \quad 
 A\ \to\ A'=1.35\,A~.
\end{equation}
It gives $m(D) = 1866\,$MeV and $m(D^*) = 2009\,$MeV.
\subsection{Baryons}\label{subse:bar}
For symmetric baryons $(m,m,m)$, the same pattern is observed as for mesons. For instance, in a simple linear potential $\sum r_{ij}/2$, the ground state is found at $E\sim 3.863$ in the NR case for $m=1$, and is lowered to $3.522$ in the SR case. This corresponds to a decrease by about 9\%, very similar to the decrease by 8\% observed for a two-body system with masses $m_1=m_2=1$ and bound by $V(r)=r$. 

Other mass combinations have been studied, with the results given in Table~\ref{tab:bar-lin}. There is a smooth transition from the SR to the NR regime. 
\begin{table}[h!]
\caption{Comparison of the SR and NR energies for a baryon with masses $m_i$ and potential $\sum r_{ij}/2$\label{tab:bar-lin}}
\begin{ruledtabular}\begin{tabular}{cccccc}
$m_1$  & $m_2$ & $m_3$  & NR & SR & Diff(\%) \\
\hline
1   & 1 & 1 & 3.863 & 3.522 & 9  \\
1   & 4 & 4 & 2.985& 2.800  & 6  \\
1  & 4  &8  &2.860 &2.671   & 7 \\
1  &10 &10 &2.644 &2.454 &7   \\
1  &10 &15 &2.591 &2.398 &7   \\
1  &10 &20 &2.561 &2.366 &7   \\
1  &20 &30 &2.430 &2.222 &8   \\
1  &30 &40 &2.353 &2.149 &8   \\
10& 30& 40	& 1.419 &1.413 & 0.5 \\
20  & 30&  40	& 1.272 & 1.270 & 0.2 \\
30 	 & 30&  40	&1.207 & 1.206 & 0.1
\end{tabular}
\end{ruledtabular}
\end{table}

We now consider a doubly-heavy baryon $(M,M,m)$ with $m=0.5$ and $M=5$ whose quarks are bound by a pairwise  interaction $\sum V(r_{ij})/2$ and the same Coulomb-plus-linear potential $V(r)=-0.4/r+0.2\,r$ as earlier for mesons. We obtain $E\simeq 0.652$ in the SR case, lower than the $0.780$ in the NR case.

If one adopts a naive diquark model, i.e., a two-step method in which one first solves for $QQ$ using the $QQ$ potential alone, and then for $QQ$-$q$ using $V(r)$ where $r$ is the distance from the light quark to the center of the diquark,%
\footnote{To be more specific, the 2-body problem is solved with masses $M$ and $M$, and then with masses $2\,M$ and $m$, i.e., the diquark mass is not renormalized.} %
one obtains $0.746$ in the NR case, while the SR scheme gives $0.473$. This means that the distortion induced by the naive diquark model is seemingly amplified in semi-relativistic calculations. 
\section{Tetraquarks}\label{se:tetra}
We now study tetraquarks, first in simple toy models, and then in more realistic potentials AL1 and AL1N tuned to reproduce the ordinary hadrons. We concentrate on two types of mass distribution for $QQ\bar q\bar q$: $MMmm$ for which binding is expected if the mass ratio is large enough, and the ``fully-heavy" configurations, $QQ\bar Q\bar Q$, for which there are somewhat conflicting predictions in the literature. 
\subsection{Doubly-flavored tetraquarks}
For $QQ\bar q\bar q$, it is known for decades that in a static and flavor-independent potential, the system becomes bound if the mass ratio $M/m$ is large enough. Moreover, in the case of $QQ\bar u\bar d$ with isospin $I=0$ , the binding is favored by the chromomagnetic interaction. See, e.g., \cite{Richard:2016eis}, and the recent discussion about various dynamical effects \cite{Richard:2018yrm}. 

\begin{figure}[b!]
 \centering
 \includegraphics[width=.35\textwidth]{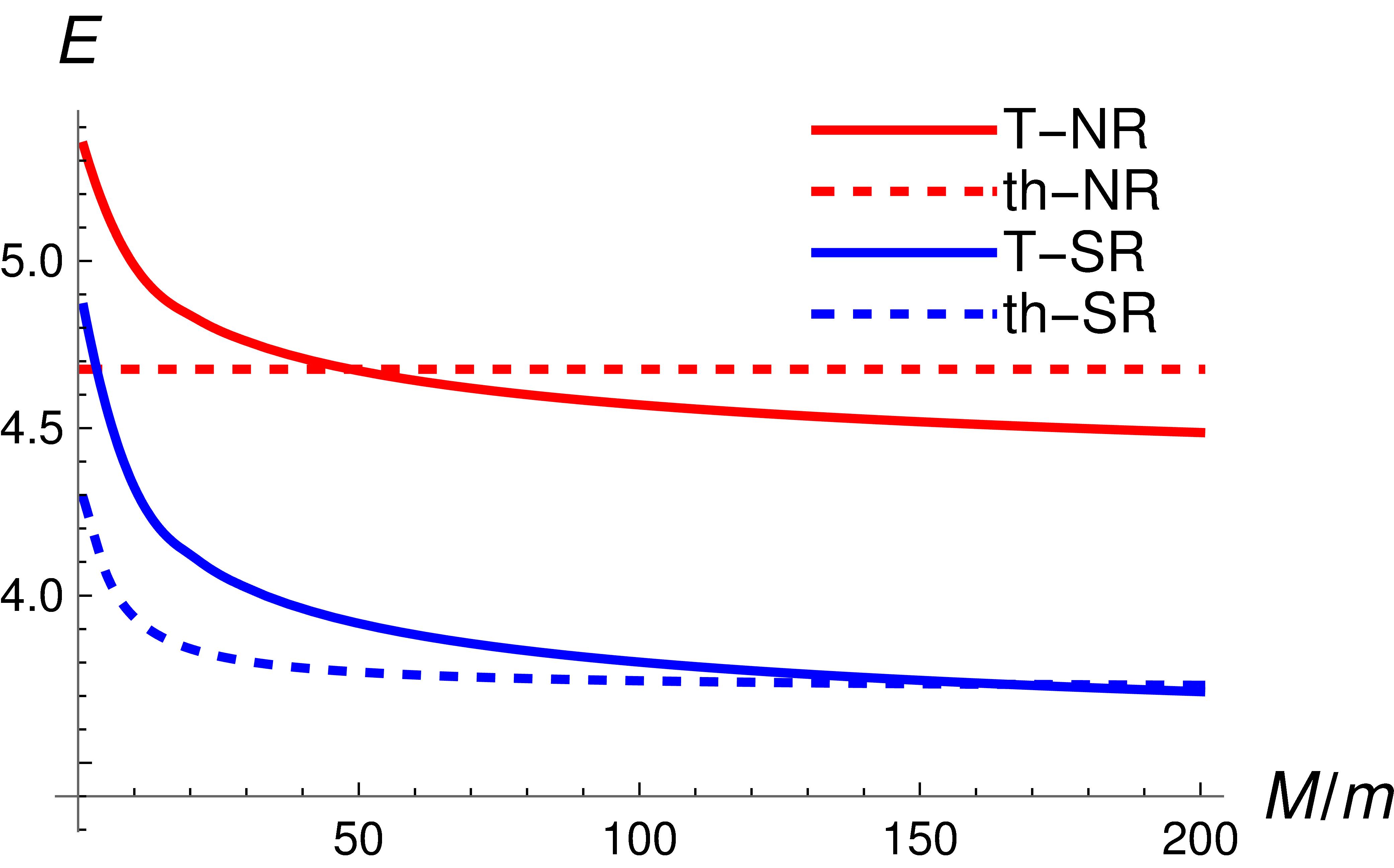}
 \caption{Comparison of the tetraquark energies with frozen color T bound by the simple linear potential \eqref{eq:V-lin}. The masses are such that $M^{-1}+m^{-1}=2$. The energy scale is fixed by the sum of inverse masses and the string tension set to unity.
\label{fig:tetra-T-lin}}
\end{figure}

\begin{figure}[h!]
 \centering
 \includegraphics[width=.35\textwidth]{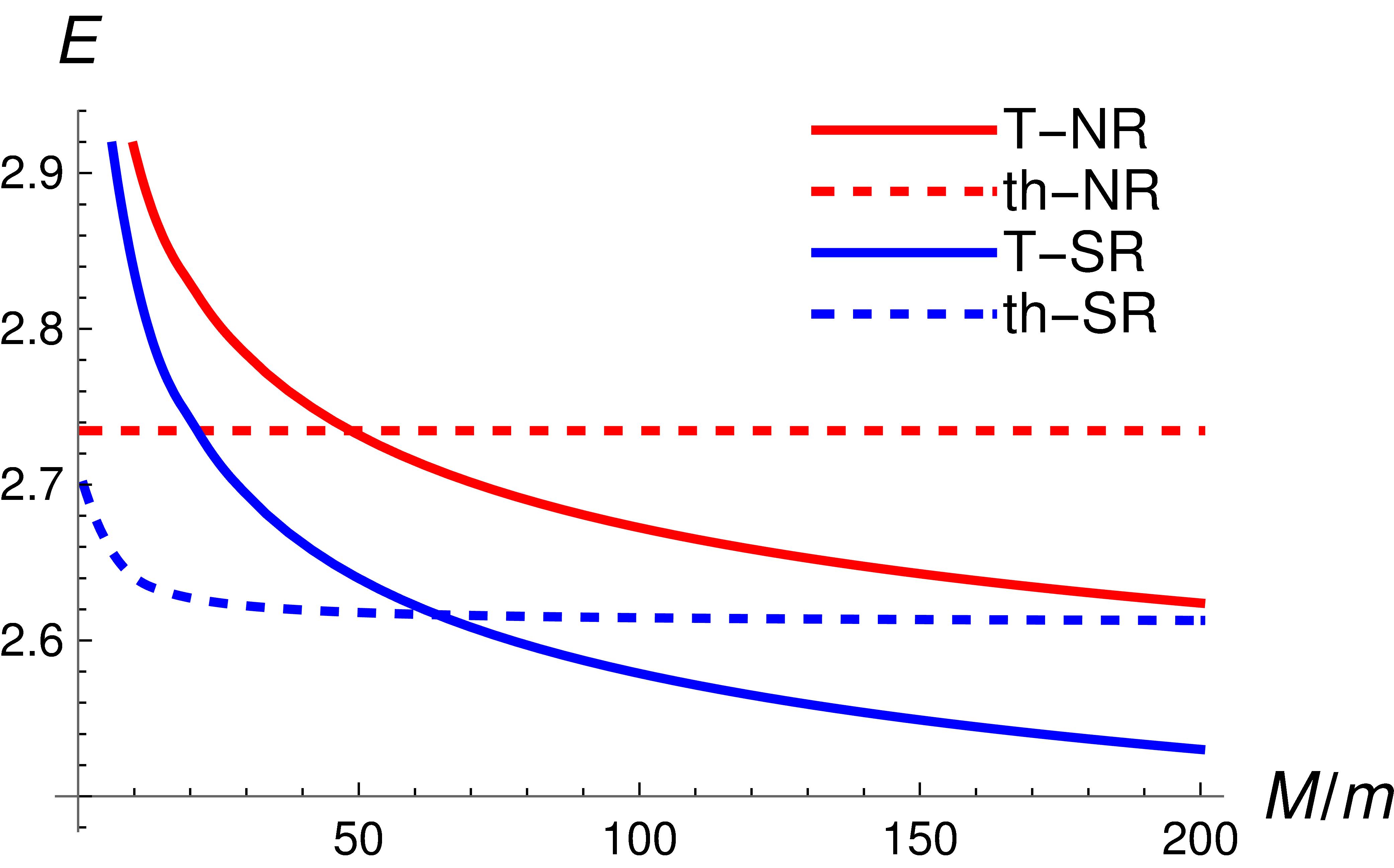}
 \caption{Same as Fig.~\ref{fig:tetra-T-lin} but for $M^{-1}+m^{-1}=0.4$.  \label{fig:tetra-T-lin-2}}
\end{figure}

We start with a purely chromoelectric model, namely a potential
\begin{equation}\label{eq:V-lin}
 V=-\frac{3}{16} \sum_{i<j}^4 \ll{i}{j} \,r_{ij}~,
\end{equation}
and utilize the usual notation $\mathrm{T}=\bar 33$ and $\mathrm{M}=6\bar6$ for the color states in the $qq$-$\bar q\bar q$ basis. 
The energy scale is fixed by imposing a potential of unit strength for a quark-antiquark pair forming a meson. In Fig.~\ref{fig:tetra-T-lin} we show the results for a T configuration of color. The mass ratio $M/m$ is varied with $M^{-1}+m^{-1}=2$ kept constant, so that the NR threshold is fixed at $E=4.6762$. It is clear that binding occurs at larger mass ratio in the semi-relativistic case. Note that at fixed $M^{-1}+m^{-1}$, the SR threshold depends slightly on $M/m$, as shown in Figs. \ref{fig:tetra-T-lin}, \ref{fig:tetra-T-lin-2}, and  \ref{fig:Coul05}.
If the exercise is repeated with larger masses $M^{-1}+m^{-1}=0.4$, the NR threshold being now at $ E= 2.7347$, see Fig.~\ref{fig:tetra-T-lin-2}, the critical mass ratios of the SR and NR cases become closer, but still, stability occurs for a smaller value of the mass ratio $M/m$ in the NR case than in the SR one. 

We have checked that with T-M mixing, still with a purely linear potential, the same pattern is observed.

\begin{figure}[h!]
 \centering
 \includegraphics[width=.35\textwidth]{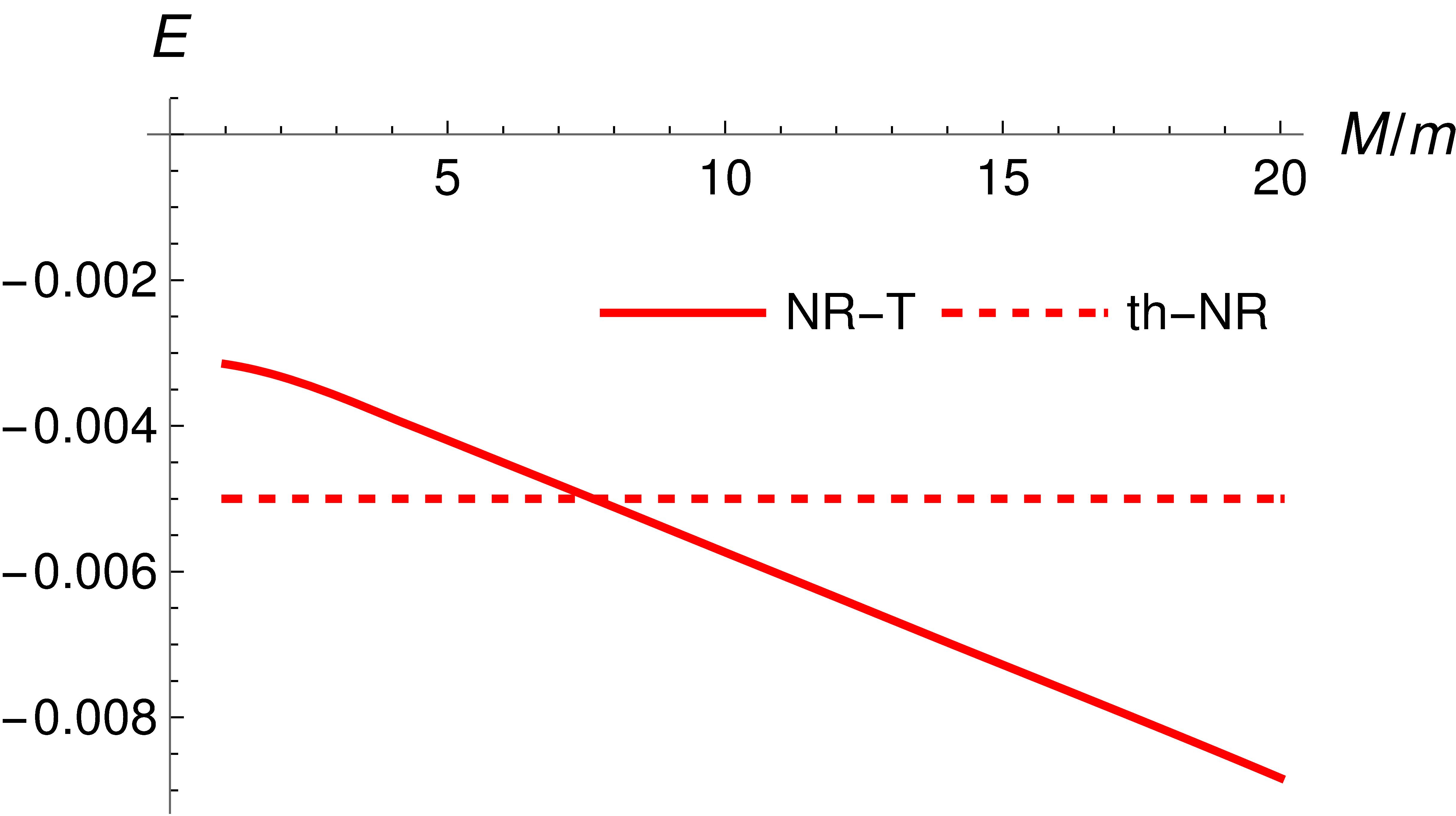}
 \caption{Ground state energy for a $QQ\bar q\bar q$ tetraquark in a T-color state with a pure Coulomb potential $-0.1/r$, compared to the threshold energy, as a function of the quark to antiquark mass ratio $M/m$, with $M^{-1}+m^{-1}=2$ kept constant. For this weak coupling, the NR and SR tetraquark and threshold energies are almost identical. The energy scale is fixed by the sum of inverse masses.}
 \label{fig:Coul01}
\end{figure}

\begin{figure}[h!]
 \centering
 \includegraphics[width=.35\textwidth]{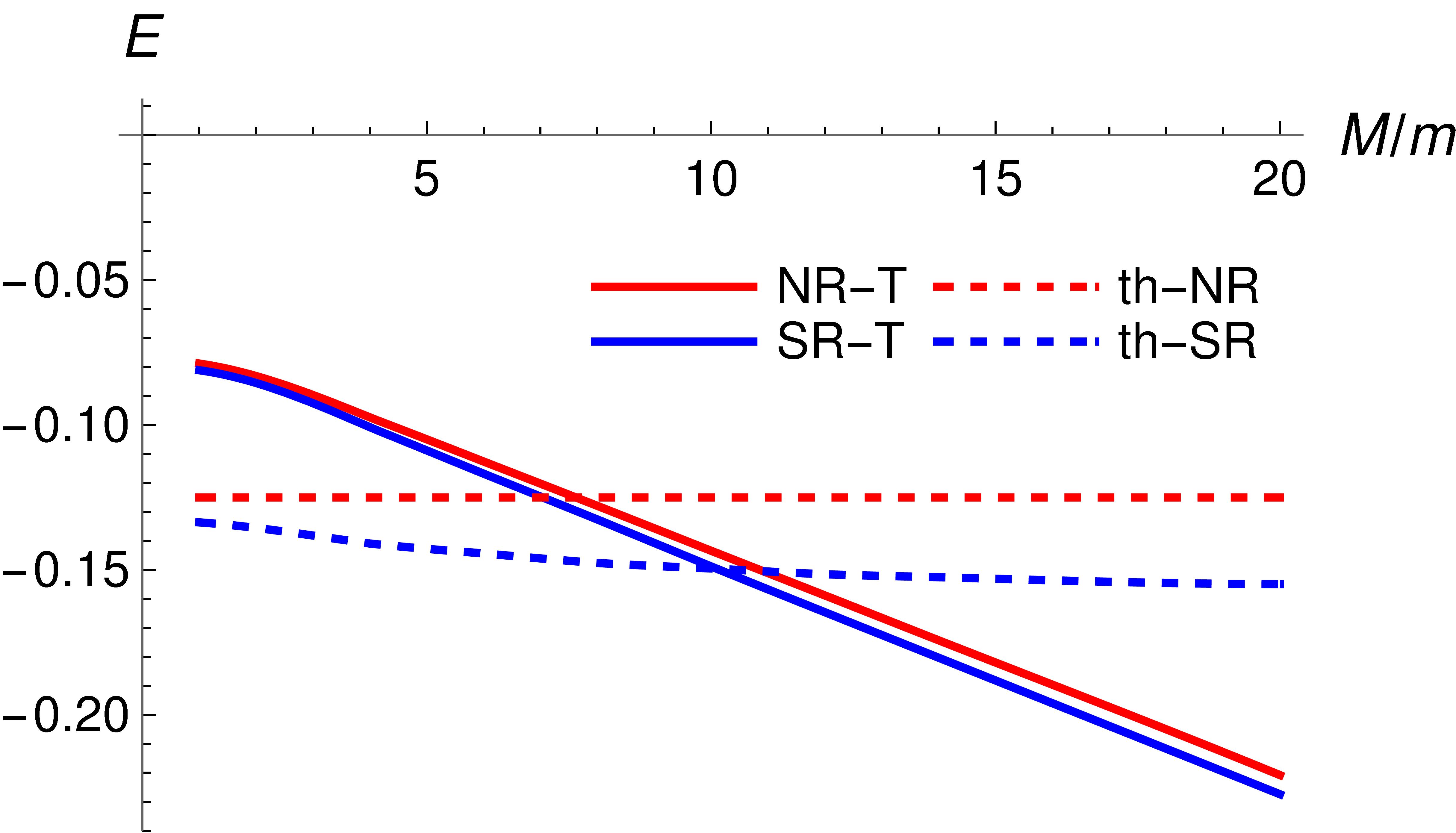}
 \caption{NR and SR energies for a T-color $QQ\bar q\bar q$ tetraquark in the limit of a pure Coulomb interaction $-0.5/r$, as  a function of the mass ratio $M/m$, with $M^{-1}+m^{-1}=2$ kept constant. The NR curves are just rescaled from Fig.~\ref{fig:Coul01}. The energy scale is fixed by the sum of inverse masses.}
 \label{fig:Coul05}
\end{figure}

We have repeated the study with a purely Coulombic interaction, that would ideally describe a bound state of very heavy quarks, provided that none of these quarks decay weakly too fast. This corresponds to the Hamiltonian 
\begin{equation}\label{eq:V-Coul}
 V=-\frac{3}{16} \sum_{i<j}^4 \ll{i}{j} \,v(r_{ij})~,
 \quad v(r)=-\frac{g}{r}~.
\end{equation}
The study is restricted to a color T. The pattern is similar to the one observed for a linear interaction, but the critical mass ratio for tetraquark binding is significantly lowered.  
For a weak coupling $g=0.1$, see Fig.~\ref{fig:Coul01}, the NR and SR energies are very close, as expected. Their comparison is just a check of the consistency of our computation scheme. 
The results corresponding to a somewhat larger coupling $g=0.5$
are shown in Fig.~\ref{fig:Coul05}.
The tetraquark energy is moderately lowered by relativistic effects, much less than the threshold energy. Hence the mass ratio required for stability is higher, and when stability is reached, the binding of the tetraquark is smaller. 

\begin{figure}[h!]
 \centering
 \includegraphics[width=.4\textwidth]{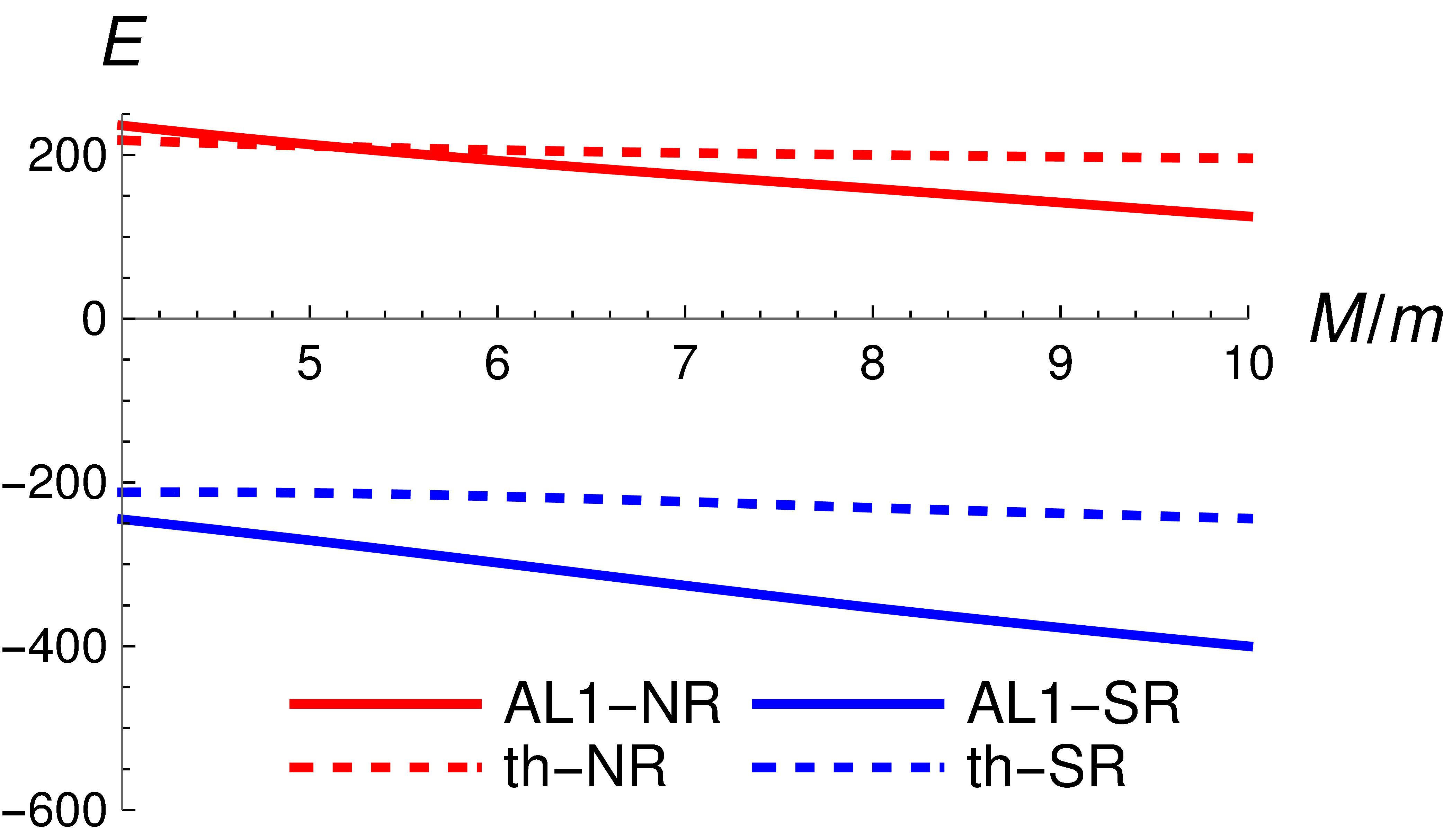}
 \caption{AL1 potential. Tetraquark vs.\ threshold for both NR and SR forms of kinetic energy. Energies in MeV.}
 \label{fig:tetra-AL1}
\end{figure}

We now do the calculation with the aforementioned AL1 model~\eqref{eq:AL1}, including the mixing of the T and M color components.
The doubly-heavy tetraquarks have been already studied by several authors with this model \cite{Janc:2004qn,Barnea:2006sd}, with the good surprise that a tiny binding is obtained for $cc\bar u\bar d$, and, of course, a more pronounced binding for the heavier analogs where one or two $c$ quarks are replaced by $b$. This is shown in Fig.~\ref{fig:tetra-AL1}: the curve labeled AL1-NR crosses the threshold th-NR before $M=m_c$, if one fixes beforehand $m=m_q$. Two remarks here are in order:
\begin{itemize}
 \item 
a value of the variational energy $E>E_{th}$ simply means that binding is not found. The lowest energy of the 4-body Hamiltonian is $E_{th}$. This is confirmed by the observation that the color content of the variational wave function tends to $1/3$ T and $2/3$ M, corresponding to a meson-meson decomposition. 
\item
the energy with a frozen M color wave function, not shown, is higher than for T, except near $M=m$, as discussed, e.g., in \cite{Richard:2018yrm}.
\end{itemize}

The SR analogs are also shown in Fig.~\ref{fig:tetra-AL1}, and the tetraquark is bound for any value of $M/m$. This is due to an unrealistic strength for the hyperfine component, which is attractive for the $\bar u\bar d$ pair, as discussed in~\eqref{eq:AL1}.

\begin{figure}[h!]
 \centering
 \includegraphics[width=.35\textwidth]{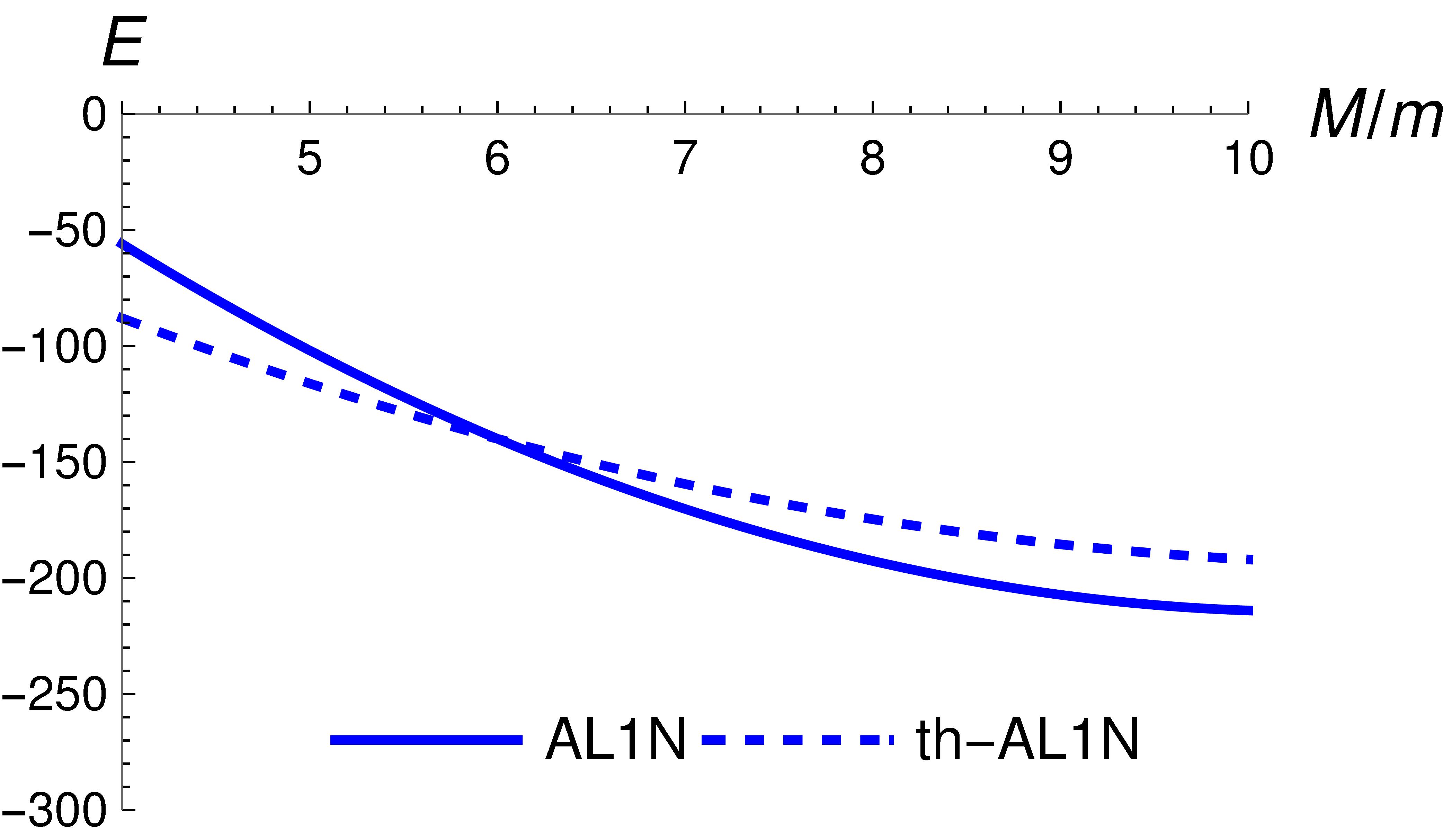}
 \caption{Modified AL1 potential. Energy of the lowest $QQ\bar q\bar q$ state calculated with relativistic kinematics, vs.\ the threshold energy. Energies in MeV.}
 \label{fig:tetra-AL1N}
\end{figure}

Using the more realistic AL1N potential, we obtain the pattern shown in Fig.~\ref{fig:tetra-AL1N} for the tetraquark vs.\ its threshold. It can be seen that once again, binding is obtained for a larger value of the mass ratio $M/m$ than in the NR case. 
\subsection{All-heavy tetraquarks}
We now consider the case of tetraquarks with two heavy quarks and two heavy antiquarks. 
There is a flurry of calculations, in particular following the announcement by LHCb of the peak in the $J/\psi$-$J/\psi$ distribution, and its interpretation as a $cc\bar c\bar c$ resonance.  
One of the questions raised by the LHCb discovery is whether there exist bound states of $cc\bar c\bar c$ or $bb\bar b\bar b$.  
Most of calculations are done in a simplified scheme with a diquark  and an antidiquark. As shown elsewhere \cite{Richard:2018yrm}, this is not a good approximation to the standard quark model,\footnote{Of course, if the diquark is introduced in the formulation of the theory, we are not dealing with an approximation but with an alternative model, subject to others pros and cons.}  and hence we do not include the corresponding papers in our discussion. 
Using a standard (color-dependent, pairwise) potential and an expansion on an harmonic-oscillator basis, Llyod and Vary \cite{Lloyd:2003yc} found a bound $cc\bar c\bar c$ bound state, but their result was not confirmed by other authors.\footnote{In our opinion, this is due to their use of an individual-particle basis and a cumbersome subtraction of the center of mass energy.}\@ In \cite{Richard:2018yrm,Richard:2019cmi} and references therein, an explanation is given on why in the chromoelectric limit $QQ\bar Q\bar Q$ is not bound, while the electric analog $e^+e^+e^-e^-$ is stable against dissociation into two $e^+e^-$ atoms.

For instance, if we adopt the above AL1 potential, the lowest $bb\bar b \bar b$ state is estimated at $18.872\,$GeV, above the threshold $18.848\,$GeV, and this energy would decrease toward this threshold if the variational expansion were pushed further. The unbound character is reinforced by the observation that the color content is nearly exactly $1/3$ for T and $2/3$ for M, which corresponds to a singlet-singlet content. The analog for SR form of kinetic energy is $18.792\,$GeV for the variational estimate, again with $33\%$ T and $67\%$ M, above the threshold  at $18.772\,$GeV. So the relativistic form of kinetic energy does not rescue the binding of $QQ\bar Q\bar Q$. 
\section{Outlook}\label{se:out}
Clearly, estimating the mass and properties of ordinary and exotic hadrons requires sophisticated tools where long-range and short-range aspects of the dynamics are well accounted for. Simple models are nevertheless useful to probe some mechanisms, provided these models are solved carefully. 

In this article, we revisited the effect of relativistic kinematics in the quark model by studying how the results are modified by replacing the NR form of kinetic operator $\vec p^2/(2\,m)$ by $\sqrt{\vec p^2+m^2}-m$. In the meson sector, most relativistic effects can be absorbed by a tuning of the parameters. Such change is illustrated in Eq. (\ref{eq:AL1N}). The readjustment of the constituent masses is not very dramatic, and could be supplemented by minor changes of the static potential. More important is the necessary modification of the chromomagnetic term, because the relativistic corrections influence the short-range correlations.

The same is true for baryons. The study becomes more delicate for tetraquarks. For a given potential, both the threshold energy and the multiquark energy are lowered by the change of kinetic energy. It is observed that the effect is more pronounced for the former, so that the binding energy decreases. In particular, for $QQ\bar q\bar q$ configurations in a given potential, the mass ratio $M/m$ at which the system becomes stable is larger for the semi-relativistic models than for the non-relativistic ones. Besides, fully-heavy tetraquarks remain above threshold if relativistic kinematics is used.

In other words, the threshold ``benefits'' more from the relativistic corrections than the collective configuration. This is similar to what is observed for some symmetry breakings. For instance, if one starts from a $QQ\bar q\bar q$ system with masses $\{M,M,m,m\}$, and breaks the particle identity, then one usually observes that the system  $\{M,M',m,m'\}$ with $M'> M$ and $m'> m$  is less bound with respect to the $\{M',m'\}+\{M,m\}$ threshold than the more symmetric system with respect to its threshold $2\,\{M,m\}$.

\acknowledgments
This work has been partially funded by Ministerio de Econom\'\i a, Industria y Competitividad
and EU FEDER under Contracts No.\ FPA2016-77177, PID2019-105439GB, and RED2018-102572-T.

 \end{document}